\documentclass[aps,prl,twocolumn,groupedaddress,showpacs,showkeys]{revtex4-1}

\usepackage{amsmath}
\usepackage{latexsym}
\usepackage{epsfig}
\usepackage{epstopdf}
\usepackage{makeidx}
\usepackage{amsfonts}
\usepackage{mathtools}
\usepackage{color}
\usepackage{float}

\begin{document}

\title{ Anomalous random networks}

\author{H. Zhang}
\email{zhanghong13@cdut.cn}
\affiliation{College of Mathematics and Physics, Chengdu University of
Technology, Cheng'du, Si'chuan 610059, China}

\author{G. H. Li}
\email{liguohua13@cdut.cn}
\affiliation{College of Mathematics and Physics, Chengdu University of
Technology, Cheng'du, Si'chuan 610059, China}

\date{\today}

\begin{abstract}
After the groundbreaking work of Erd$\ddot{o}$s-R$\acute{e}$nyi random graph, the random networks has made great progress in recent years. One of the eye-catching modeling is
time-varying random network model capable of encoding the instantaneous time description of the network dynamics. To further describe the random duration time for the nodes to be inactive, we herein propose a dinner party anomalous random networks model,
and derive the analytical solution of the probability density function for the node being active at a given time.  Moreover, we investigate the gift delivery and viral transmission in dinner party random networks. This work provides new quantitative insights in describing random networks, and could help model other uncertainty phenomena in real networks.

\end{abstract}


\pacs{89.75.Hc, 05.60.-k}

\maketitle

\section{Introduction}

Network modeling and the related dynamic mechanism has got great attention in recent decades. A Lot of the models use random networks to describe
the uncertainty in real-world networks. The early random model
appeared in a series of papers starting in 1959 by Erd$\ddot{o}$s and  R$\acute{e}$nyi. They investigated what a ‘typical’
graph which arises by taking $n$ vertices, and placing an edge between
any pair of distinct vertices with some fixed probability $a$. \cite{H2016,FM2008}
 Motivated by Erd$\ddot{o}$s-R$\acute{e}$nyi random model there are some other extensions to capture
 randomness and different characteristics in nature networks. \cite{ER2002,BAJ1999}
 
 Recently, to describe
the instantaneous and fluctuating dynamics of many networks, Perra et al \cite{PBMGPV2012,PGPV2012} propose a time-varying random networks by defining the
activity potential, a time invariant function characterizing the agents' interactions and constructing an
activity driven model capable of encoding the instantaneous time description of the network dynamics. This
model provides an explanation of structural features such as the presence of hubs, which originate
from the heterogeneous activity of agents.

Note that in present random network models, the  duration time for the nodes to be active is often exponential distributed, which corresponds to the constant inactive rate. However, for real-world complex networks, the distribution for the active times is non-exponential, and obeys power-law distribution. 
We shall herein develop a dinner party anomalous random networks model to address this question based on the renewal theory. The dynamical processes of this random network with arbitrary active waiting times are
described analytically. Moreover, the evolution of gift delivery and viral transmission, which respectively represents diffusion process and  reaction process, in dinner party anomalous random networks are also discussed.

\section{Renewal process and the renewal equation}

Let $X_1, X_2,...$ be a sequence of nonnegative, independent, identically distributed random variables with distribution function
$P(X_i\leq t)=F(t)$. We let $S_0=0$.
Define for $t>0$ the renewal process $N(t)=\sup\{n\in N: S_n\leq t\}$ with $N(t)=0$ if $X_1>t$, where $S_n$ are the sum $S_n=X_1+...+X_n, n\geq 1$ with $P(S_n\leq t)=F_n(t)$.
The main attention in renewal theory is given to the renewal function
$M(t)=E[N(t)]$ and is given by  $M(t)=\sum_{n=1}^{\infty} F_n(t)$ where $F_n(t)=\int_{0}^{t}F_{n-1}(t-u)dF(u)$ and $F_1(t)=F(t)$.
It is well-known that $M(t)$ is the solution of the renewal equation
\begin{eqnarray}
 M(t)&=&F(t)+\int_{0}^{t}M(t-t')dF(t')\nonumber\\
 &&=F(t)+\int_{0}^{t}F(t-t')dM(t').
\end{eqnarray}
The renewal density satisfies the equation
\begin{eqnarray}
 h(t)&=&f(t)+\int_{0}^{t}h(t-t')f(t')dt'\nonumber\\
 &&=f(t)+\int_{0}^{t}f(t-t')h(t')dt'.
\end{eqnarray}
where $h(t)dt=dM(t)$ denotes the renewal probability in $[t,t+dt]$  and $f(t)=F'(t)$ which is the PDF for waiting time $t$ to renew.

\section{Time varying random networks and anomalous random network}

\subsection{Time varying random networks}

In this section we start by recalling the time varying random networks  generated according to the following memoryless stochastic process \cite{PBMGPV2012}:
(i) At each time step $\Delta t$, the instantaneous network $G_t$ starts
with $N$ disconnected nodes. (ii) With probability $a_i\Delta t$,
each vertex $i$ becomes active and generates $m$ undirected
links that are connected to $m$ other randomly selected
vertices, where $a_i$ is a data extracted from a distribution
$F(a)$. Nonactive nodes can still receive connections
from other active vertices. (iii) At time $t+\Delta t$, all the
edges in the network $G_t$ are deleted and the process starts
over again to generate the network $G_{t+\Delta t}$. It can be shown
that the full dynamics of the network is encoded in the
activity rate distribution $F(a)$.

\subsection{Anomalous random network and its activity renewal rate}
Inspired by the above model, we herein propose a dinner party anomalous random networks model in which each node waits for some random time to contact several friends to have a dinner party. This random networks satisfies the following rules:
 (i) At initial time the  network $G_t$ starts
 with $N$ nodes. (ii) For each node it waits for a random time $\tau_1$ to  become active and generate $m$ undirected
 links that are connected to other $m$ randomly selected
 vertices. Nonactive nodes can receive connections
 from other active vertices. (iii) At time $\tau_1+\Delta \tau$, the
 edges of node $i$ are deleted, and then the process is renew for node $i$, that is, the node $i$ has to wait for another time $\tau_2$ to hold a new party). Here the party duration time $\Delta \tau$ is too transitory comparing to the waiting time $\tau_i$ and is ignored for simplicity.
If at time $t$ one node is not only active such that it can connect $m$ links  but also can be connected by the other active node, then the two parties will be organized together.
 We assume the renew processes of all nodes are independent, and at the initial time the nodes can not be active. We can see that for each node $i$ the activation process
 $N_i(t)=\sup\{n\in N: S_{in}\leq t\}$ where $S_{in}$ are the sum $S_{in}=\tau_{i1}+...+\tau_{in}, n\geq 1$ with $P(S_{in}\leq t)=F_{in}(t)$ and $F_{i1}(t)=F_i(t)$
 is a renewal process. We denote the renewal function  by $M_i(t)=E[N_i(t)]$. Let $p_i(t)dt=dM_{i}(t)$. Then
 $p_i(t)dt$ represents the activity probability of the node $i$ holding a party in the time interval $[t,t+dt]$. Let $\psi_i(t)=F'_{i}(t)$.
So $\psi_{i}(t)$ is the probability density function (PDF) for a node $i$ waiting for time $t$ to begin a new party, and
 $\Psi_{i}(t)=1-F_i(t)$ is the probability not holding the party in $(0,t)$.
According to the renewal density equation,
we obtain
\begin{eqnarray}
 p_{i}(t)=\psi_{i}(t)+\int_{0}^{t}p_{i}(t')\psi_{i}(t-t')dt',
\end{eqnarray}
for $t>0$.

Taking Laplace transform of Eq.(1) yields the result:
\begin{eqnarray}
 p_{i}(s)=\frac{\psi_{i}(s)}{1-\psi_{i}(s)},
\end{eqnarray}
where $p_{i}(s)$ and $\psi_{i}(s)$ are the Laplace transforms of $P_{i}(t)$ and $\psi_{i}(t)$.

\subsection{One example}
We then consider the first special cases of $p_i(t)$. Noticing that for exponential distribution
\begin{eqnarray}
\psi_i(t)=a_i e^{-a_i  t},
\end{eqnarray}
where $a_i$ for node $i$ is a data extracted from a distribution $F(a)$.
Eq.(2) reduce to  $p_{i}(s)=\frac{a_i }{s}$, whose inverse Laplace transform is $p_{i}(t)=a_i$, and thus $p_{i}(t)dt=a_i dt$.

 This means that in this case the probability $p_{i}(t)dt$ of the node $i$ being active in time interval $[t,t+dt]$ does not depend on $t$, which recovers the activity rate for the time varying random networks in Ref.\cite{PBMGPV2012}.

 \subsection {The other example}
If we consider the power law waiting time PDF $\psi_i(t)=\frac{\alpha_i}{t+\tau_0}(\frac{\tau_0}{t+\tau_0})^{\alpha_i}$ for $0<\tau_0, 0<\alpha_i$,\cite{FS2017,SS1974}  and
whose Laplace transform is $\psi_i(s)\sim 1-\Gamma(1-\alpha_i)\tau^{\alpha_i}s^{\alpha_i}$ for small $\tau_0$, then $p_{i}(s)=\frac{\psi_{i}(s)}{1-\psi_{i}(s)}=\frac{1}{\Gamma(1-\alpha_i)\tau_0^{\alpha_i}}s^{-\alpha_i}$, by taking the inverse laplace transform we finds $p_i(t)=\frac{1}{\Gamma(\alpha_i)\Gamma(1-\alpha_i)\tau_0^{\alpha_i}}t^{\alpha_i-1}$. Here, $\alpha_i$ is assigned according to a given probability
distribution $F(\alpha)$.

 One can see that in this case the activation  rate $\frac{1}{\alpha_i\Gamma(\alpha_i)\Gamma(1-\alpha_i)\tau_0^{\alpha_i}}[(t+dt)^{\alpha_i}-t^{\alpha_i}]$ in $[t,t+dt]$ which depends on the time $t$. Therefore, the activation rate of the node $i$ has memory on how long the time has passed away.
 We call such a random network an anomalous time-varying random network.

 \subsection{Degree}

 Next we want to know how many directed (undirected) links there are for node $i$ in $[t,t+dt]$. Firstly, we focus on the directed linked event. Since in our assumption once $i$ is active, it will link $m$ nodes randomly, and will have degree $m$. Thus, the average degree of $i$ in $[t, t+dt]$ is $m p_i(t)dt$ by considering the activation probability $p_i(t)dt$.

 Next we turn to the undirected case, in which the linked event from $i$ to the other node $j$ depends not only on the activation event of node $i$ and the event that $i$ links to $j$ but also on the activation event of node $j$ and the event that $j$ links to $i$.  This linked event from $i$ to $j$ can be divided into five incompatible parts as following:

(i) The node $i$ is active, $j$ is not active, and $i$ links $j$.

(ii) The node $j$ is active, $i$ is not active, and $j$ links $i$.

(iii) The nodes $i$ and $j$ are both active, $i$ links $j$, but $j$ does not link $i$.

(iv) The nodes $i$ and $j$ are both active, $j$ links $i$, but $i$ does not link $j$.

(v) The nodes $i$ and $j$ are both active, $j$ links $i$, and $i$ links $j$.

Therefore, the probability $p_{i\leftrightarrow j}(t)dt$ of undirected event from $i$ to $j$ in $[t,t+dt]$ is the sum of the probabilities of the five events, that is,
 \begin{eqnarray}
p_{i\leftrightarrow j}(t)dt&=&p_{i}(t)dt(1-p_{j}(t)dt)\frac{m}{N-1}\nonumber\\
&&+(1-p_{i}(t)dt)p_{j}(t)dt\frac{m}{N-1}\nonumber\\
&&+p_{i}(t)dt p_{j}(t)dt\frac{m}{N-1}(1-\frac{m}{N-1})\nonumber\\
&&+p_{i}(t)dt p_{j}(t)dt (1-\frac{m}{N-1}dt)\frac{m}{N-1}\nonumber\\
&&+p_{i}(t)dt p_{j}(t)dt\frac{m}{N-1}\frac{m}{N-1}\nonumber\\
&&=(p_{i}(t)dt+p_{j}(t)dt)\frac{m}{N-1}\nonumber\\
&&-p_{i}(t)dt p_{j}(t)dt(\frac{m}{N-1})^2.
 \end{eqnarray}
Note also that in the events (i)-(iv), there is one undirected link between $i$ and $j$ and in (v) there is two. Next we want to  know the degree of node $i$  in $[t, t+dt]$ (that is, the numbers of links from and to $i$) under the conditions of the five cases (i)-(v), respectively. We find that under (i) and (iii), the degrees $k_{i,1}$ and $k_{i,3}$ both equal to $m+\frac{m}{N-1}\sum_{k\neq i,j}p_k(t)dt$, where $m$ is the number of links from the active $i$ and $\frac{m}{N-1}\sum_{k\neq i,j}p_k(t)dt$ is the number of links to $i$ from other active nodes; under (ii), the degree is $k_{i,2}=1+\frac{m}{N-1}\sum_{k\neq i,j}p_k(t)dt$ where $1$ is the number of links from the active $j$ and $\frac{m}{N-1}\sum_{k\neq i,j}p_k(t)dt$ is the number of links directed to $i$ from other active nodes except $i$ and $j$; and under (iv) and (v), the degrees $k_{i,4}$ and $k_{i,5}$ are both $m+1+\frac{m}{N-1}\sum_{k\neq i,j}p_k(t)dt$, where $m$ is the number of links from the active $i$, $1$ is the number of links from the active $j$ and $\frac{m}{N-1}\sum_{k\neq i,j}p_k(t)dt$ is the number of links directed to $i$ from other active nodes except $i$ and $j$.

\section{Diffusion and reaction in anomalous random networks }

\subsection{Diffusion in anomalous random networks}
In this section we will consider a gift delivery process (or random walk for one particle) in the dinner party random networks.
It is assumed that for the gift in the node $i$, once the node $i$ is active, the gift immediately diffuses away with probability $1$, to one of its neighbors with the uniform distribution, while if the node is not active the gift will keep staying at node $i$.
  We also assume that in one party  a gift must transfer only for once from the inviter  to an  invitee. We now want to know where the sole gift is at time $t$.
Let $\rho_{i}(t)$ be the PDF for the gift being at the node $i$ at time $t$, and let $\rho_{i}^{-}(t)$ and $\rho_{i}^{+}(t)$ be respectively loss and gain PDFs of the gift at $i$  at time $t$.
 Then one has
 \begin{equation}
 \rho_{i}(t)=\rho_{i}(0)\Psi_{i}(t)+\int_{0}^{t}\rho_i^{+}(\tau)\Psi_{i}(t-\tau) d\tau
 \end{equation}
 which means that  the gift is at $i$ in the initial time $0$ and does not go away, or the gift arrives at $i$ at some time $0< t-\tau< t$  and does not go away in $[\tau, t]$.

Let $\rho_{i}^{-}(t)$ be the loss probability for the gift at the node $i$ at time $t$. Then we have
\begin{equation}
 \rho_{i}^{-}(t)=\rho_{i}(0)\psi_{i}(t)+\int_{0}^{t}\rho_i^{+}(\tau)\psi_{i}(t-\tau) d\tau
 \end{equation}
 which means that  the gift is at $i$ in the initial time $0$ and goes away at time $t$ or the gift arrives at $i$ at some time $0< t-\tau< t$  and goes away at $t$.

 Taking laplace transforms of Eqs. (7) and (8) respectively, we find
  \begin{eqnarray}
\rho_{i}(s)=[\rho_i (0)+\rho_i^{+}(s)]\Psi_i(s),
  \end{eqnarray}
   and
  \begin{eqnarray}
 \rho_{i}(s)=[\rho_i (0)+\rho_i^{+}(s)]\psi_i(s)
  \end{eqnarray}

 Combining the above two equations, we find the relation between $\rho_{i}^{-}(s)$ and $\rho_{i}(s)$ as following
 \begin{eqnarray}
\frac{\rho_{i}^{-}(s)}{\rho_{i}(s)}=\Theta_i(s).
 \end{eqnarray}
 Here,  $\Theta_{i}(s)=\frac{\psi_i(s)}{\Psi_i(s)}$.
 
For exponential waiting time $\psi_i(t)=a_i e^{-a_i  t}$, when $\Psi_i(t)=e^{-a_i  t}, \Theta_i(t) = a_i\delta(t)$ \cite{ADH2013,SSS2006}, the above relation becomes
 \begin{eqnarray}
\rho_{i}^{-}(s)=a_i\rho_{i}(t).
 \end{eqnarray}

Noting that the gain probability for the gift at the node $i$ at time $t$ comes from the sum of the lost probabilities of the gift in the other nodes, one gets
$\rho_{i}^{+}(t)=\sum_{j\neq i}\rho_{j}^{-}(t)\frac{m}{N-1}\frac{1}{m}$ where $\frac{m}{N-1}=p_{i\leftrightarrow j,2}(t)+p_{i\leftrightarrow j,4}(t)+p_{i\leftrightarrow j,5}(t)$ is the PDF for the node $j$ holding a party and inviting the node $i$, and multiplied the terms $\rho_{j}^{-}(t)$ and $\frac{1}{m}$ to represent at time $t$ the gift in $j$ goes away to the node $i$ from a random directed link.
Thus, we can
obtain the master equation for the evolution of $\rho_{i}(t)$ for exponential waiting time as following
 \begin{eqnarray}
\frac{d\rho_{i}(t)}{dt}&=&\rho_{i}^{+}(t)-\rho_{i}^{-}(t)\nonumber\\
 &&=\sum_{j\neq i}a_j\rho_{j}(t)\frac{1}{N-1}-a_i\rho_{i}(t).
\end{eqnarray}

\subsection{Reaction in anomalous random networks}
In this case we shall investigate viral transmission in dinner party anomalous random networks.
We assume that
 in one party the viral  may transfer from one node to every node in the party  with the probability $p$. The reaction process are as following:
  \begin{eqnarray}
I+S\to I+I; I\to R.
\end{eqnarray}
Here, the symbols $I$, $s$ and $R$ represents the infected, susceptible, and recovery nodes, respectively. The susceptible node is infected because of contacting  infected node in their party. So when he is infected depends on the active waiting time of parties, on if he holds or takes part in one party where someone else is already infected.
 
 We  want to know the infected probability of each node at time $t$ and how many nodes has been infected at time $t$.
 Let $I_{i}(t)$ be the infected probability of the node $i$ at time $t$. Then the whole number of infected nodes is $I(t)=\sum_{i} I_{i}(t).$ Assume that $\psi_{r}(t)$ is the recovery PDF for an infected node $i$ waiting for time $t$ to recover and $\Psi_{r}(t)=1-\int_{0}^{t}\psi_{r}(t')dt'$ is the  probability for the infected node being ill in $[0,t]$.  Then we have
 \begin{equation}
 I_{i}(t)=I_{i}(0)\Psi_{r}(t)+\int_{0}^{t}I_{i}^{+}(t')\Psi_r (t-t')dt',
 \end{equation}
where $I_{i}^{+}(t)$ be  adding infected probability of the node $i$ at time $t$. One can find
 \begin{equation}
 I_{i}^{+}(t)=p\sum_{j} \tilde{p}_{i\leftrightarrow j}(t)I_{j}(t).
 \end{equation}
where $\tilde{p}_{i\leftrightarrow j}(t)$ is the PDF for two nodes $i$ and $j$ meeting in a party at time $t$, satisfying
 \begin{eqnarray}
\tilde{p}_{i\leftrightarrow j}(t)&=& p_{i\leftrightarrow j}(t)+\sum_{k\neq i, j}p_{k}(t)\frac{C_{N-3}^{m-2}}{C_{N-1}^{m}}\nonumber\\
&&=(p_{i}(t)+p_{j}(t))\frac{m}{N-1}-p_{i}(t)p_{j}(t)(\frac{m}{N-1})^2\nonumber\\
&&+\sum_{k\neq i, j}p_{k}(t)\frac{m(m-1)}{(n-1)(n-2)},
 \end{eqnarray}
If the waiting time PDFs for all nodes holding parties are same and equalling to $\psi_i(t)=\frac{\alpha}{t+\tau_0}(\frac{\tau_0}{t+\tau_0})^{\alpha}$ where $0<\alpha<1$ is a random variable with probability
distribution $F(\alpha)$ which leads to $p_i(t)=\frac{1}{\Gamma(\alpha)\Gamma(1-\alpha)\tau_0^{\alpha}}t^{\alpha-1}$,  then the contacting PDF between two nodes $i$ and $j$ becomes
  \begin{eqnarray}
\tilde{p}_{i\leftrightarrow j}(t)&=& \frac{1}{\Gamma(\alpha)\Gamma(1-\alpha)\tau_0^{\alpha}}\bigg[t^{\alpha-1}\frac{m(m-1)}{N-1}\nonumber\\
&&-\frac{1}{\Gamma(\alpha)\Gamma(1-\alpha)\tau_0^{\alpha}}  t^{2\alpha-2}(\frac{m}{N-1})^2 \bigg].
 \end{eqnarray}

 Denoting the loss probability (just the recovered probability) of node $i$ at time $t$ by $ I_{i}^{-}(t)$,
 we have
\begin{equation}
 I_{i}^{-}(t)=I(0)\psi_{r}(t)+\int_{0}^{t}I_i^{+}(\tau)\psi_{r}(t-\tau)d\tau.
 \end{equation}
Then we can get the number of recovered
 nodes $R(t)$ until the time $t$ as following:
 \begin{equation}
 R(t)=\sum_{i}\int_{0}^{t}I_{i}^{-}(t')dt'.
 \end{equation}
 Thus, the number $S(t)$ of susceptible nodes is $S(t)=N-R(t)-I(t)$.

From the Laplace transforms of Eqs. (15) and (19) we can get $I_{i}^{-}(t)=\int_{0}^{t}I_{i}(t')\theta_{r} (t-t')dt'$, where $\theta_{r}(t)$ is the inverse Laplace transform of $\theta_{r}(s)=\frac{\psi_r (s)}{\Psi_r(s)}$, and $\psi_r (s), \Psi_r (s)$ are the Laplace transform of $\psi_r (t), \Psi_r (t)$.
We now can get the master equation for the evolution of $I_{i}(t)$ as following
 \begin{eqnarray}
\frac{dI_{i}(t)}{dt}&=&I_{i}^{+}(t)-I_{i}^{-}(t)\nonumber\\
 &=&p\sum_{j}\langle\tilde{p}_{i\leftrightarrow j}(t)\rangle I_{j}(t)-\int_{0}^{t}I_{i}(t')\theta_r(t-t')dt'.~~~
 \end{eqnarray}
Here, $\langle\tilde{p}_{i\leftrightarrow j}(t)\rangle$ is the mean of $\tilde{p}_{i\leftrightarrow j}(t)$ with respect to the distribution $F(\alpha)$, that is,
$$\langle\tilde{p}_{i\leftrightarrow j}(t)\rangle=\int_{0}^{1}\tilde{p}_{i\leftrightarrow j}(t)dF(\alpha).$$

As an example, we finally analyse the epidemic spreading with power law recovery waiting time whose PDF is given by
$\psi_r(t)=\frac{\beta}{t+\tau_0}(\frac{\tau_0}{t+\tau_0})^{\beta}$ for $0<\tau_0, 0<\beta$.\cite{FS2017,SS1974}
In Laplace space one has $\psi_r(s)\sim 1-\Gamma(1-\beta)\tau^{\beta}s^{\beta}$ for small $\tau_0$,
 and thus $\theta_r (s)=\frac{1}{\Gamma(1-\beta)\tau_0^{\beta}}s^{1-\beta}$. Inserting it into the Laplace transform of Eq.(21) and taking the inverse Laplace transform, we then obtain
the equation describing the time evolution for $I_i(t)$
 \begin{eqnarray}
\frac{dI_{i}(t)}{dt}&=&p\sum_{j}\langle\tilde{p}_{i\leftrightarrow j}(t)\rangle I_{j}(t)\nonumber\\
 &&-\frac{1}{\Gamma(1-\beta)\tau_0^{\beta}}(_0D_t^{1-\beta}I_{i}(t)),
 \end{eqnarray}
where $_0D_t^{1-\beta} f(t)$ is the  Riemann-Liouville fractional derivative operator, equalling in Laplace $t\rightarrow s$ space to
$s^{1-\beta} \textit{L} [f(t)]$. \cite{MR1993} Because of the effect of the fractional kinetics operator the epidemic spreading with power law waiting time in anomalous random networks has a memory dependence on the infected probability of the node at previous
times.

\section{Conclusion}

In summary, we introduce the
dinner party anomalous random networks model  to capture the effect of the random inactive duration time in random networks, and derive the solution (3) for the node $i$ being active at a given time.  Moreover, we discuss the gift delivery in such a dinner party random networks model, and investigate the
effect of the random active duration time of each node on viral transmission in dinner party anomalous random networks. This work provides new quantitative insights in describing random networks, and also provide a suited theoretical approach for modeling other  uncertainty phenomena in real networks.
\newline

\section{acknowledgments}
 This work was supported by the Natural Science Foundation of Sichuan Province (Grant
No. 2022NSFSC1752).\newline

\section{Appendix A}
\begin{eqnarray}
 h(t)&=&(M(t))'=(\sum_{n=1}^{\infty} F_{n}(t))'=\sum_{n=1}^{\infty} p_{n}(t)\nonumber\\
 &&=f(t)+\sum_{n=2}^{\infty} p_{n}(t)\nonumber\\
&&=f(t)+\sum_{n=2}^{\infty}\int_0^t f(t-t') p_{n-1}(t')dt'\nonumber\\
&&=f(t)+\int_0^t f(t-t')d(\sum_{n=1}^{\infty} F_{n}(t'))\nonumber\\
&&=f(t)+\int_0^t f(t-t')h(t')dt'.
\end{eqnarray}
where $p_{n}(t)dt=dF_{n}(t)$ and $p_{1}(t)=f(t)$.

\end{document}